\documentstyle[11pt,paspconf,epsf]{article}

\begin{document}

\title{Using Numerical Simulations to Gain Insight into the
Structure of Superbubbles}
\author{Philip T. Komljenovic\altaffilmark{1,2}, Shantanu Basu, and Doug Johnstone}
\affil{Canadian Institute for Theoretical Astrophysics, 60 Saint George St.,
Toronto, ON, M5S 3H8, Canada}

\altaffiltext{1}{Also: Department of Physics and Astronomy,
York University, Toronto, ON, M3J 1P3, Canada}
\altaffiltext{2}{Present address: Department of Physics and Astronomy,
University of Victoria, Victoria, BC, V8W 3P6, Canada}

\newcommand{\HI}{H{\small I}~}
\newcommand{\HII}{H{\small II}~}

\begin{abstract}
Recent high resolution observations of Galactic superbubbles have 
motivated us to re-examine several classes of superbubble models.
We compare three classes of hydrodynamic models (the Kompaneets
approximation, the thin shell model, and numerical simulations) in
order to understand the structure of superbubbles and to gain
insight into observations. In particular, we apply models to
the W4 superbubble, which has been observed in the
Pilot project of the arcminute resolution Canadian Galactic Plane Survey
(Normandeau et al. 1996).
Magnetohydrodynamic simulations are also performed and point the
way to a fuller understanding of the W4 superbubble. We suggest that
the highly collimated bubble and apparent lack of a Rayleigh-Taylor
instability in the superbubble shell can be explained by the presence
of a magnetic field. 
\end{abstract}

\keywords{hydrodynamics - ISM:bubbles - MHD - shock waves }

\section{Introduction}

In a recent paper, Basu, Johnstone, \& Martin (1999) have modeled the
shape and ionization structure of the W4 superbubble 
(see Normandeau \& Basu, these proceedings) using the  
semianalytic Kompaneets (1960) model for blast wave propagation in 
a stratified exponential atmosphere. Our motivation in this paper
is to compare the simplified Kompaneets model with more sophisticated
models for superbubble expansion in a stratified medium. 
Our study reveals the differences between the various models, but also
shows that the more detailed hydrodynamic models face difficulties
in properly accounting for the shape of the W4 superbubble. However,
the highly collimated W4 superbubble may be fit
by numerical models which include a magnetic field with a significant 
vertical (perpendicular to the Galactic plane) component.

\section{Comparison of Hydrodynamic Models}

\begin{figure}
\vspace{13cm}
\includegraphics{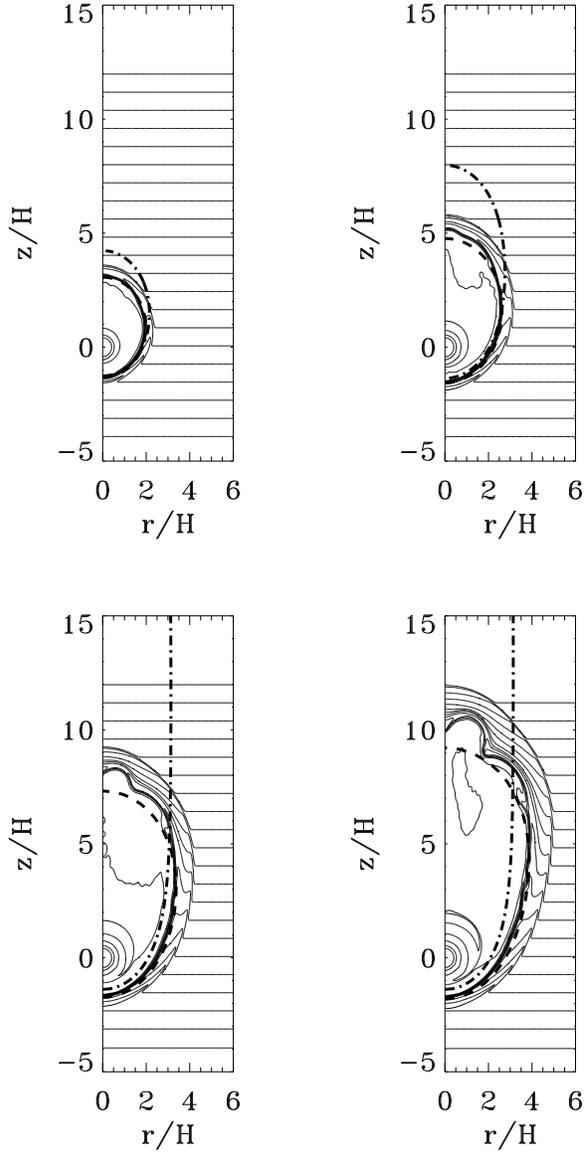}
\caption{A comparison of numerical simulations (solid density contours;
logarithmically spaced)
with the position of the shell in the Kompaneets (dash-dotted line)
and thin shell approximation (dashed line), for an exponential atmosphere
$\rho(z) = \rho_0 \exp(-z/H)$. The four snapshots occur at
dimensionless times 4.36, 6.36, 8.37, and 9.33. The unit of time
is $[t] = (\rho_0 H^5/L_0)^{1/3}$, where $L_0$ is the wind luminosity.}
\end{figure}

We compare three classes of models for superbubble expansion
in stratified media. The earliest one is due to Kompaneets (1960), 
and describes the propagation of a strong shock wave
in an exponentially stratified medium. An analytic solution exists
for the bubble shape at different times, and the time evolution is
obtained by solution of an ordinary differential equation.
The thin shell approximation (MacLow \& McCray 1988) assumes a geometrically
thin shell and determines its motion by direct integration of the
momentum equation for various segments of the shell.
Finally, numerical simulations (e.g., MacLow, McCray, \& Norman 1989)
provide the most complete hydrodynamic solutions. We reproduce all 
three calculations for an adiabatic bubble in an exponential atmosphere.
The numerical simulations are performed with the ZEUS-2D code.

Figure 1 compares the shape of the bubble at four different times for
the three cases. The solid density contours represent the numerical
solution. The Kompaneets model (dash-dotted line) expands most rapidly
and assumes a more elongated shape than the other models. Its more rapid
evolution means that it has already blown out of the atmosphere in the 
two lower panels. The thin shell approximation (dashed line) remains 
close to the inner boundary of the shell of swept-up mass in the numerical
solution. The closeness of the thin shell and numerical results occurs
since the thin shell model tracks the motion of the swept up mass.
The numerical solution
is also prone to a Rayleigh-Taylor instability at late times, when the
upper shell is accelerating rapidly. We note that the Rayleigh-Taylor 
instability would be more pronounced if cooling was allowed in the 
shell (MacLow et al. 1989). An important difference between the models
is that the Kompaneets model is more highly collimated before blowout
than the other two models, since the vertical acceleration 
is unhindered by inertial effects, and occurs very rapidly.

\section{Magnetohydrodynamic Model}

Due to the inertial effects mentioned above, the numerical hydrodynamic 
and thin shell results 
cannot produce highly collimated bubbles that meet the aspect ratio of the
W4 superbubble, even though the less realistic Kompaneets model can do so.
From H$\alpha$ observations of the ionized shell (Dennison, Topasna, \&
Simonetti 1997), we measure an aspect ratio 
$\frac{z_{\rm top}}{r_{\rm max}}\simeq 3.3$,
where $z_{\rm top}$ is the distance from the star cluster to the top of the
bubble, and  $r_{\rm max}$ is the maximum half width of the bubble.
We have attempted a variety of atmospheric models in the numerical
and thin shell models, and find that neither a steeper nor shallower
profile than the exponential stratification can explain the high 
collimation of the W4 superbubble.

We address this issue by carrying out magnetohydrodynamic (MHD) simulations.
A vertical (along $z$) magnetic field provides the necessary external
pressure at large heights to confine the bubble to a narrow width.
Figure 2 shows the evolution of the bubble in an exponential 
atmosphere with an initial
vertical magnetic field $B_z = 3$ $\mu$G. At the latter time, the
aspect ratio of the bubble matches the observed aspect ratio of the
H$\alpha$ shell.

\begin{figure}
\vspace{8cm}
\includegraphics{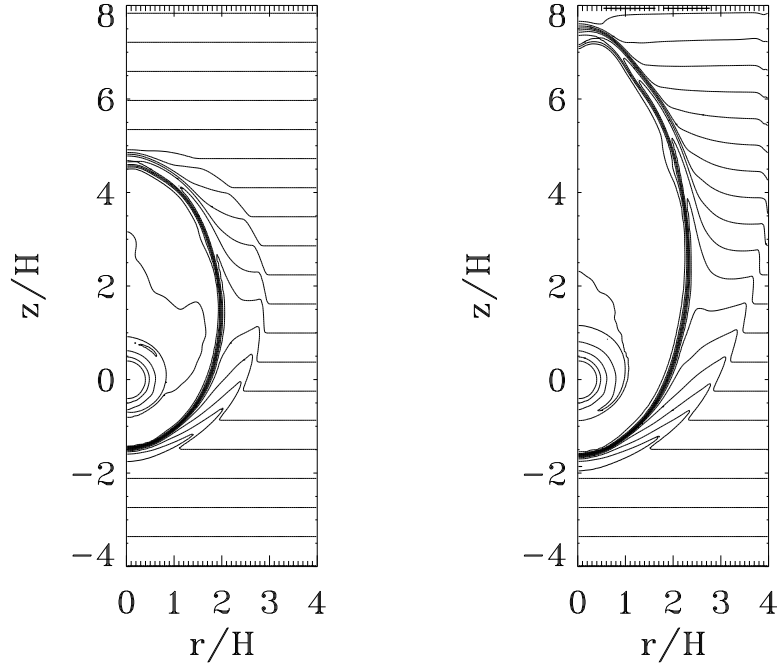}
\caption{MHD model. Density contours of the expanding superbubble in
an exponential atmosphere with an initially vertical magnetic field
$B_z = 3$ $\mu$G. Dimensionless times are 5.35 and 6.77, where the unit
of time is the same as in Fig. 1.}
\end{figure}

\section{Discussion}

Although the Kompaneets model can provide a highly collimated bubble
that matches the aspect ratio of the W4 superbubble, more realistic
hydrodynamic models predict wider bubbles than observed in W4, due
to their proper accounting of inertial effects. However, a numerical
model which includes a vertical magnetic field of a few $\mu$G strength
can achieve the required collimation.
We believe that such a field can also suppress the potential
Rayleigh-Taylor instability in the accelerating upper shell, thereby
explaining why the observed H$\alpha$ shell of the W4 superbubble 
does not appear to be breaking up.
Although a purely vertical field is an idealization and is not supported
by Faraday rotation data, we point out that even an initially horizontal
magnetic field can produce a significant vertical component through
the action of the Parker instability (e.g., Basu, Mouschovias, \& Paleologou
1997). More realistic magnetic 
field geometries such as these remain to be explored.



\begin{references}
\reference Basu, S., Johnstone, D., \& Martin, P. G. 1999, ApJ, in press 
\reference Basu, S., Mouschovias, T. Ch., \& Paleologou, E. V. 1997, ApJ, 480,
L55
\reference Dennison, B., Topasna, G. A., \& Simonetti, J. H. 1997, \apj, 474, L31
\reference Kompaneets, A. S. 1960, Sov. Phys. Dokl., 5, 46
\reference MacLow, M.-M., \& McCray, R. 1988, ApJ, 324, 776
\reference MacLow, M.-M., McCray, R, \& Norman, M. L. 1989, ApJ, 337, 141
\reference Normandeau, M., Taylor, A. R., \&  Dewdney, P. E. 1996, Nature, 380, 687

\end{references}
\end{document}